\documentclass{aa}

\begin{document}
\newcommand{\up}[1]{\ifmmode^{\rm #1}\else$^{\rm #1}$\fi}
\newcommand{\zdot}{\makebox[0pt][l]{.}}
\newcommand{\upd}{\up{d}}
\newcommand{\uph}{\up{h}}
\newcommand{\upm}{\up{m}}  
\newcommand{\ups}{\up{s}}
\newcommand{\arcd}{\ifmmode^{\circ}\else$^{\circ}$\fi}
\newcommand{\arcm}{\ifmmode{'}\else$'$\fi}
\newcommand{\arcs}{\ifmmode{''}\else$''$\fi}

\title{A Catalog of OB Associations
in the spiral galaxy NGC 300 \thanks{Based in part on  observations obtained 
with the 1.3~m
Warsaw telescope at the Las Campanas  Observatory of the Carnegie
Institution of Washington, and on observations obtained at the European Southern 
Observatory, La Silla, Chile}}

\author{G. Pietrzy\'nski
      \inst{1}\fnmsep\thanks{On leave from Warsaw University Observatory}
  W. Gieren
      \inst{1}
P. Fouqu\'e
      \inst{2}\fnmsep\inst{3}
F. Pont
     \inst{4}
}

\institute{Universidad de Concepci{\'o}n, Departamento de Fisica,
 Casilla 160--C, Concepci{\'o}n, Chile\\
\email{pietrzyn@hubble.cfm.udec.cl, wgieren@coma.cfm.udec.cl}
\and
Observatoire de Paris-Meudon DESPA, F-92195 Meudon CEDEX, France
\and
European Southern Observatory, 
Casilla 19001, Santiago 19, Chile\\
\email{pfouque@eso.org}
\and
Universidad de Chile, Departamento de Astronomia, Casilla 36D, Santiago, Chile\\
\email{FredericPont@obs.unige.ch}} 

\date{}

\abstract{We present results of a search for OB associations  in  NGC 300. 
Using an automatic and objective method (PLC technique) 117 objects were found. 
Statistical 
tests indicate that our sample is contaminated by less than 10 
detections due to random concentrations of blue stars. Spatial distributions 
of detected associations and H II regions are strongly correlated.
The size distribution reveals a significant peak at about 60 $\mu$rad which 
corresponds 
to 125 parsecs if a distance modulus of 26.66 mag is assumed. Besides the objects 
with sizes corresponding to typical associations we also found several 
much larger objects. A second level application of our detection method revealed 
that most of these are composed of smaller subgroups, with sizes of about 100 
pc.
}
\maketitle


\section{Introduction}
OB associations constitute physical groups of young, massive, gravitationally
unbound stars, which have formed in the same molecular cloud.
Observations of these short lived 
objects  provide much valuable information  about regions of recent and/or 
current
star formation, parent galaxies and star formation
processes in general. 
In particular, a comparison of the properties 
of systems of associations in different  galaxies may help us to better
understand the influence of the environment on mechanisms of 
star formation. 

A lot of work has been dedicated in the past to the detection and 
analysis of properties of associations in nearby galaxies. However,
as thoroughly discussed by Hodge (1986), the identification of 
OB associations is a very difficult task. Indeed, due to very low surface
density of stars it is not possible to detect them just based on
the density contrast with the surrounding field. For a long time, associations 
were identified from visual searches of photographic plates. 
Subjective identification together with nonuniform observational material 
and poor resolution often led to disagreement between different authors.
As a result of these limitations it was not possible to perform unambiguous
comparisons of the properties of stellar associations found in different
galaxies. To overcome these difficulties new, automatic and objective techniques 
of 
detecting OB associations were developed (Battinelli 1991, Wilson 1991).
All one needs in order to use these methods are the positions of the sample of 
OB stars.
A successful application of these techniques to search for OB associations
in several nearby galaxies (Battinelli 1991, Magnier et al. 1993, Wilson 1992) 
reflects that they are very useful
tools for building uniform catalogues of OB associations in galaxies well suited 
for
comparison purposes. 

In this contribution we present the results of a search for OB
associations in the nearly face-on spiral galaxy NGC 300. This galaxy, which is 
a member
of the Sculptor Group, presents massive recent star formation, yet no systematic 
work
to discover OB associations in this stellar system has been undertaken.
 To our knowledge only five stellar groups assumed to be associations were 
observed   by Breysacher et al. (1997) during their search for WR stars in stellar associations.
 A systematic search
with the aim to uncover the complete system of OB associations in this important 
nearby galaxy is therefore long overdue.

Our paper is organized as follows. In Sect. 2 we briefly describe the 
observations and 
data reduction techniques applied in this work. The algorithm used for searching 
for 
potential associations is described in Sect. 3. Results are presented 
in Sect. 4. Sect. 5 contains some final remarks on our findings.

\section{Observations}
The observational data have been collected in the course of a long term
project whose principal goal is to discover a large number of Cepheid variables 
in NGC 300 in order to calibrate the effect of metallicity 
on Cepheid luminosities.
 Up to now
several hundred images in $BVRI$ bands were collected with the Wide Field Camera 
attached to the ESO/MPI
2.2 m telescope on La Silla. This camera has a pixel scale of 0.238 arcsec/pixel,
and its large field of view of 33 x 34 arcmin allows to cover the whole galaxy 
with one pointing. 
 For the search of OB associations we used 
 10 $B$ and 10 $V$ images obtained  on two photometric nights, at low
airmass and under 0.8 arcsec seeing conditions. Exposure time was set to 
360 sec for all images in each filter. As can be seen from Fig. 1 the limiting 
magnitude was about 22.5 mag in $B$ and about 23 mag in $V$.
After debiasing and flatfielding the images with the IRAF \footnote{IRAF is 
distributed by the 
National Optical Astronomy Observatories, which are operated by the 
Association of Universities for Research in Astronomy, Inc., under cooperative
agreement with the NSF.}   package, each of the eight 
2k x 4k chips was divided into 16 slightly overlapping sub-frames. Profile 
photometry with $DAOPHOT $ and $ALLSTAR$ programs was performed for all stars 
detected in all subframes. Then the photometry on all subframes was tied 
together
using stars located in the overlapping regions.

In order to calibrate our data we performed follow--up observations with
the  Warsaw 1.3 m, photometric telescope located at Las Campanas
Observatory, Chile. The telescope was equipped with a 2048 x 2048 CCD camera. 
The scale is 0.41 arcsec/pixel, which corresponds to a field of 
view of about 14.5 x 14.5 arcmin.  More details about the instrumental system
 can be found in Udalski et al. (1997).
Observations were conducted through 
$BVRI$ filters, during 2 photometric nights. We monitored four fields, covering 
about 25 x 25 arcmin of the central part of NGC 300. During each
night a large number of Landolt standard stars (Landolt 1992) covering a wide
range of colors and airmasses,  were also observed in order
to tie our new secondary standards in NGC 300 to the Landolt system.
 Preliminary reductions (i.e. debiasing and
flatfielding) were performed with the IRAF package while the PSF photometry 
was done with the $DAOPHOT$/$ALLSTAR$ package. 

The accuracy of the  transformation of the instrumental magnitudes and colors to 
the standard system is about 0.04  
mag for the $B$ and $V$ bands, and slightly better for the $(B-V)$ color. 
As a next step we carefully selected objects  with reliable photometry and used 
them to calibrate
the photometry from the WFI camera. In this process, we eliminated background 
galaxies, severely blended and saturated stars by visual examination.

Having established a sequence of about 300 photometric secondary standard stars 
in the
field of  NGC 300 (results to be published in a forthcoming paper), we 
transformed the WFI photometry from each chip using a least
squares method. Usually more than 20 
stars common between a given chip and our catalog of secondary standards 
were identified, and the accuracy of internal transformations was
better than  0.01 mag (rms). 

We are continuing the observations of secondary standards stars in NGC 300, and 
the accuracy of their brightness 
calibration will
 soon be significantly improved. A more detailed description of the project,
 observations, and reduction procedures will be presented in a followup paper.

\section{Search for Associations}
\subsection{Identification  Method} 
The  Path Linkage Criterion (PLC) technique (Battinelli
1991) was adopted in our search for OB associations in NGC 300.
For a detailed description of this technique the reader is referred 
to Battinelli (1991). In short,
this method assumes that any given two stars in an ensemble of OB stars belong to 
the same association if and only if it is
possible to connect these two stars, by successively linking OB stars located 
between
them, separated from each other by no more than a certain fixed distance 
parameter, or search radius, 
called $d_{\rm s}$. The advantage of the PLC method is that  $d_{\rm s}$ can 
be unambiguously derived based on the given catalog of early
type stars using the function $f_{\rm p} (d)$. This function describes the number of
groups containing at least p stars, for any given value of the distance parameter 
d.
The behaviour of this function is as follows: a steep rise until a maximum 
value is reached, followed by a gradual decrease towards a limiting value of 1, 
which corresponds 
to the situation that the value of d is so large that all stars can be connected using it
(i.e. all stars are assigned to just one group). The optimum value of the 
distance parameter $d_s$ is defined as the 
value of d corresponding to the maximum of the function $f_{\rm p}(d)$.  

\subsection{Application of the PLC technique to NGC 300}

Before we can start to search for OB associations we need to select the catalog 
of blue stars. 
In the case of the photometric data it means that we need to adopt brightness 
and color cutoffs. The sample of blue stars was selected from the photometric 
catalog 
of stars in NGC 300 using the following criteria: $V<$ 22 mag and -0.6  $< B-V <$ 0.4 
mag. We note that
NGC 300 has a very low foreground reddening, in the order of $E(B-V)=0.02$ mag, so 
the color cutoff for the
unreddened $B-V$ color is also approximately 0.4 mag.
Altogether 4016 stars satisfied our selection criteria and entered our list of 
blue stars.
 Unfortunately the  distance to 
NGC 300 is rather uncertain at the present time;  distance moduli obtained by 
different authors differ quite significantly. 
Based on an analysis of a variety of distance indicators, van den Bergh (1992) 
found a distance 
modulus of 26.0 mag, while Freedman et al. (1992) obtained a value of 26.66 mag, 
using $BVRI$ CCD observations of Cepheids in this galaxy. 
If we adopt the latter value, our brightness cutoff corresponds to an absolute 
magnitude of $M_{v}=-4.7$ mag.
In Fig. 1 we present the $B-V$ color - magnitude diagram (CMD) from about 14000 
stars observed 
in the field of NGC 300, with an indication of the location of the selected blue 
stars.

The next step was to derive the value of $d_{s}$ based on our catalog of blue 
stars. To do this we apply 
the PLC technique with several values for the distance parameter and several 
values for the minimum number of stars
to define an association, p. The results are presented in Fig. 2. It can be seen 
that the position of the maximum of
the function $f_{p}(d)$ depends slightly on p and is located in the interval  
from  about 35   
to  40 microradians.
In order to check on the influence of our choice for the value of $d_{s}$ on the 
number of associations we find, we 
used three search distances, namely $d_{\rm s}$= 35, 37.5 and 40 microradians which 
correspond to linear distances of 
72, 78 and 83 pc, respectively ( assuming a distance modulus of 26.66 mag), and 
then compared the
results. We found that all but three objects were common in all trials. This is 
because in most 
cases we can just add a few (10 - 15 \%) new stars to already detected 
associations 
if we use a slightly larger value of $d_{\rm s}$. As a conclusion, the number of 
detected OB associations is very
insensitive to the adopted value of the search radius, in the range suggested by 
Fig. 2. We adopted $d_{\rm s} = 37.5 $ 
microradians as the most appropriate search radius.

The last parameter that needs to be specified is the  minimum number of stars p any 
potential association should possess. It is clear that too small a 
number of stars could produce many spurious detections.  On the other hand, if
we demand an unreasonably large number of stars many small association may
be lost. We are then at the tradeoff between completeness and
reliability of our sample of associations. 
In order to check how many detections  are caused by chance coincidence, as a 
function of p,
a statistical test was performed.
One hundred  random distributions of a number of stars equal to the number of
our sample of blue stars, and distributed over the same area, were created and 
the
PLC technique with our adopted distance parameter value  was applied to search 
for 
potential groups. This experiment was repeated for different minimum numbers 
of stars. The plot of the resulting fraction of the spurious detections to the 
total number of detected 
associations, as a function of the minimum
number of stars, is presented in Fig. 3. We can see that the spurious
detections constitute less than 10 \% if we adopt 6 stars as a minimum 
population of OB stars
of potential associations. This number seems to be a good
compromise and we therefore adopted it.

\section{Results}
\subsection{The Catalog}

The application of the PLC technique, with a threshold of number of stars equal 
to 6 and the
distance parameter value of 37.5 microradians as discussed above,  resulted in 
the detection of 117 groups of blue
stars. Table 1 contains their description. Columns 1, 2 and 3 give the 
designation, 
and the equatorial coordinates of the detected associations. The number of blue 
stars belonging to the group, its size 
in both microradians and parsecs (distance modulus of 26.66 assumed), 
 cross-identification with the catalog of HII regions in NGC 300 (Deharveng
et al. 1988) and objects studied by Breysacher et al. (1997), (arabic and roman 
numbers respectively)   are given 
in the following columns.  

The coordinates of the OB associations were determined as the mean coordinates 
of their separate member
stars. Following Battinelli (1991), the real diameter of associations is defined 
as
$0.5*(\Phi_{R.A.}*cos(\Phi_{Dec})+ \Phi_{Dec})*d_{\rm NGC~300}$, where
$\Phi_{R.A.}$ and  $\Phi_{Dec}$
 correspond to the  diameter in right ascention and declination respectively and $d_{NGC~300}$
 stands for distance to the galaxy. We assumed $d_{\rm NGC~300} 
= 2.1  Mpc$ (distance modulus of  26.66 mag).
Fig. 4 displays the spatial distribution of the detected associations over the 
disc of NGC 300. Maps of six typical associations are presented in the appendix.

 We would like to stress that our spatial resolution allows detection of
associations as small as  about 15 pc. We therefore expect that our catalog is 
near complete even down to potentially very compact associations.

\subsection{Size distribution}

 In Fig. 5 we show the size distribution of the detected associations. There 
is a sharp peak 
located at about 60 microradians.  
 This
value depends somewhat on the assumed value of $d_{\rm s}$.
 Using   $d_{\rm s} = 35$ and $d_{\rm s} = 40$ microradians, respectively, 
sizes are about 15 \% different (smaller and bigger, respectively), but 
the shape of the distribution remains the same.
Table 2 gives numerical values for the location of the peak of the size 
distribution, for a 
range of distance parameters, and for the two different distance moduli to NGC 
300 discussed before.

Until a more precise distance to NGC 300 is known (which is one of the main 
purposes of our NGC 300 project), it is not 
possible to carry out an unambiguous comparison of our size results with those 
obtained for OB associations in other galaxies.

\setcounter{table}{1}

 \begin{table}
      \caption[]{Peak of size distribution}
         \label{tabpeak}
\begin{tabular}{cccc}
\hline
\noalign{\smallskip}
peak location & $d_{\rm s}$=35 & $d_{\rm s}$=37.5 & $d_{\rm s}$=40 \\
\hline
$\mu$rad & 53 & 60 & 69 \\
in pc for $\mu =26.0$  mag & 84 & 95 & 110\\
in pc for $\mu =26.66$ mag) & 110 & 125 & 144\\
\hline
\end{tabular}
\end{table}

\subsection{Correlation with HII regions}
In most cases the ionization of hydrogen is caused by UV photons emitted from 
nearby 
hot, massive stars,  so one can suspect a correlation of the spatial 
distributions of 
OB associations and H II regions. To check this we compared the spatial 
distribution 
of OB associations obtained in this paper with the spatial distribution of H II 
regions in NGC 300 cataloged by Deharveng et al. (1988). The result is that 
about  60 \% of the detected OB associations do  clearly overlap with H II regions
 (see Tables 1 and 3 for detailed cross-identification) , which 
indicates that there is indeed the expected strong correlation. Thus both OB 
associations
and H II regions appear to be about equally well suited tracers of recent or 
ongoing star formation.

\subsection{Stellar Complexes}
We can see from Fig. 5 that besides the stellar groups 
with sizes typical for OB associations
(of about 100 pc) there are also some groups having much bigger dimensions. Regarding 
their size they resemble Galactic star complexes, as described by Efremov et al. 
(1987).

By visual inspection of the objects identified as having sizes larger than 300 
pc (potential star
complexes) we found that most of them appear to consist of a number of smaller, 
well defined
associations with sizes close to 100 pc (see the map of one of the stellar 
complexes presented in the appendix). These were not detected as individual 
stellar groupings by the PLC technique  
due to the high density
of blue stars with respect to the surrounding regions in these parts of the 
galaxy.

To check on this in an objective way we selected  blue stars from each of these 
regions
and applied the PLC method for a second time. In 14 out of 18 cases we detected 
that the
apparently large single associations do indeed consist of several smaller 
associations. Their
description is given in Table 3. The names of new objects were designated using 
the name of the corresponding associations detected during the first application 
of the PLC technique,
 followed by consecutive roman letters. (e.g. AS\_102a, AS\_102b, etc.)
Their parameters were derived in the same manner as described in Sect. 4.1. As 
a conclusion,
very large single OB associations  (like AS\_008, AS\_031, AS\_084 and 
AS\_096)  seem to be very rare.

 It is interesting to note that 14 out of 18 potential star complexes are located 
in the western part of NGC 300, while the distribution of 
detected OB associations seems to be fairly uniform (see Fig. 4). This fact suggests that
for some unknown reasons the physical  conditions 
for the formation of large objects  have been more favorable in this part
of the galaxy.

\section{Summary}
We have presented the results of a search for OB associations in the spiral 
galaxy NGC 300, using 
photometric data obtained within a broader project aiming at a precise distance 
determination to this galaxy from its Cepheid variables. Magnitudes and colors 
were obtained for 
about 14000 stars in a 0.5 x 0.5 square degree field centered on the nucleus of 
NGC 300.
From this sample, we selected blue stars using the constraints 
$V < 22$ mag and $-0.6 < B-V < 0.4$ mag. Application of the PLC method resulted in the
detection of 117 groups of blue stars. Statistical considerations show that the 
expected
contamination of this sample by spurious detections is less than 10 \%. This 
suggests that our sample 
constitutes a homogeneous group which is well suited for comparison with systems 
of associations found in other galaxies. We detected the expected strong 
correlation 
between the spatial distributions of associations and H II regions in NGC 300, a 
fact
which lends additional strong support to the relability of our OB association 
searching procedure.

Most of the detected OB associations have sizes comparable to typical 
associations from the
Galaxy and other Local Group galaxies. However, a few objects have sizes up to 
five  
times larger, corresponding rather to stellar complexes than associations. 
The PLC technique was applied once more to stars from the regions of each of the 
potential 
stellar complexes. This revealed that most of these large complexes are composed 
of 
well-defined smaller subgroups having sizes corresponding to typical OB 
associations. 
This suggests a hierarchical arrangement of OB stars groupings, already 
found previously in other galaxies (Efremov et al. 1987, Battinelli et al. 1996). 

Once we will have an accurate distance determination to NGC 300 from our 
project,
we will use the results of this paper to compare the system of OB associations 
of
NGC 300 to those of other galaxies for which such studies have been performed.

\begin{acknowledgements}
We would like to thank the OGLE team, especially Drs: Andrzej Udalski, 
Michal Szyma{\'n}ski and Marcin Kubiak for their kind help in observations 
and data reduction. We are very grateful to the European Southern Observatory 
for large amounts of observing time
with the Wide Field Camera on the 2.2 m telescope. It is a pleasure to thank the 
2p2 team for expert support
and for doing several of the observations in service mode.
 It is also a pleasure to express our gratitude to 
Dr. Paolo Battinelli for providing us with his computer programs. WG 
acknowledges financial support through Fondecyt grant No. 1000330.
 We would also like to thank Dr. Azzopardi for valuable comments.
\end{acknowledgements}

\setcounter{table}{0}

\onecolumn

\begin{table}

\caption[]{OB association candidates in  NGC 300}
\begin{tabular}{ccccccc}
\hline
Name  & $\alpha_{2000}$ & $\delta_{2000}$ & N &size &size & cross- \\ 
      &                 &                 &   &[$\mu rad$]&[pc]& identification \\
\hline
AS\_001 & 0\uph54\upm10\zdot\ups9 &  -37\arcd37\arcm43\arcs &    6 &    48 &   100 & 3 \\ 
AS\_002 & 0\uph54\upm17\zdot\ups4 &  -37\arcd35\arcm05\arcs  &   66 &   264 &   552& *\\ 
AS\_003 & 0\uph54\upm23\zdot\ups5 &  -37\arcd39\arcm53\arcs  &    9 &    78 &   162& 14\\
AS\_004 & 0\uph54\upm23\zdot\ups6 &  -37\arcd39\arcm18\arcs  &   12 &    74 &   154& \\ 
AS\_005 & 0\uph54\upm23\zdot\ups7 &  -37\arcd34\arcm44\arcs  &   15 &   126 &   263& 12,11,16\\
AS\_006 & 0\uph54\upm24\zdot\ups4 &  -37\arcd41\arcm28\arcs  &   10 &    99 &   207& \\ 
AS\_007 & 0\uph54\upm24\zdot\ups4 &  -37\arcd38\arcm16\arcs  &    8 &    82 &   172& \\ 
AS\_008 & 0\uph54\upm25\zdot\ups1 &  -37\arcd35\arcm15\arcs  &   29 &   264 &   552& \\ 
AS\_009 & 0\uph54\upm25\zdot\ups5 &  -37\arcd39\arcm34\arcs  &   50 &   208 &   435& *\\ 
AS\_010 & 0\uph54\upm25\zdot\ups9 &  -37\arcd38\arcm06\arcs  &   14 &    68 &   143& \\ 
AS\_011 & 0\uph54\upm27\zdot\ups0 &  -37\arcd43\arcm27\arcs  &    6 &    45 &    94& 19\\
AS\_012 & 0\uph54\upm27\zdot\ups9 &  -37\arcd39\arcm19\arcs  &   14 &    90 &   188& \\ 
AS\_013 & 0\uph54\upm28\zdot\ups1 &  -37\arcd37\arcm44\arcs  &    9 &    55 &   115& 22\\ 
AS\_014 & 0\uph54\upm28\zdot\ups5 &  -37\arcd41\arcm32\arcs  &   32 &   148 &   308& *\\ 
AS\_015 & 0\uph54\upm30\zdot\ups5 &  -37\arcd43\arcm47\arcs  &    9 &    41 &    87& 27\\
AS\_016 & 0\uph54\upm31\zdot\ups1 &  -37\arcd37\arcm52\arcs  &   37 &   163 &   341& *\\
AS\_017 & 0\uph54\upm31\zdot\ups8 &  -37\arcd44\arcm12\arcs  &   11 &    82 &   172& 28,33\\
AS\_018 & 0\uph54\upm31\zdot\ups9 &  -37\arcd38\arcm40\arcs  &   40 &   201 &   420& *\\
AS\_019 & 0\uph54\upm32\zdot\ups9 &  -37\arcd39\arcm11\arcs  &    7 &    52 &   108& \\ 
AS\_020 & 0\uph54\upm34\zdot\ups2 &  -37\arcd37\arcm40\arcs  &    6 &    79 &   164& \\ 
AS\_021 & 0\uph54\upm34\zdot\ups3 &  -37\arcd45\arcm48\arcs  &    6 &    68 &   142& 36\\ 
AS\_022 & 0\uph54\upm34\zdot\ups4 &  -37\arcd42\arcm54\arcs  &    7 &    47 &    98& \\ 
AS\_023 & 0\uph54\upm35\zdot\ups2 &  -37\arcd39\arcm40\arcs  &    7 &    41 &    86& 37,38\\ 
AS\_024 & 0\uph54\upm35\zdot\ups9 &  -37\arcd41\arcm08\arcs  &    8 &    74 &   156& \\ 
AS\_025 & 0\uph54\upm38\zdot\ups8 &  -37\arcd41\arcm36\arcs  &   10 &    97 &   204& 39,41\\ 
AS\_026 & 0\uph54\upm39\zdot\ups7 &  -37\arcd42\arcm45\arcs  &   34 &   197 &   411& *\\
AS\_027 & 0\uph54\upm39\zdot\ups8 &  -37\arcd40\arcm21\arcs  &   37 &   150 &   314& *\\
AS\_028 & 0\uph54\upm40\zdot\ups1 &  -37\arcd39\arcm27\arcs  &    6 &    55 &   116& \\ 
AS\_029 & 0\uph54\upm41\zdot\ups1 &  -37\arcd40\arcm54\arcs  &   42 &   219 &   459& *\\ 
AS\_030 & 0\uph54\upm41\zdot\ups4 &  -37\arcd35\arcm46\arcs  &   11 &    88 &   184& 47\\ 
AS\_031 & 0\uph54\upm41\zdot\ups7 &  -37\arcd41\arcm48\arcs  &   24 &   193 &   404& \\ 
AS\_032 & 0\uph54\upm41\zdot\ups9 &  -37\arcd38\arcm33\arcs  &    6 &    68 &   143& 48\\
AS\_033 & 0\uph54\upm42\zdot\ups1 &  -37\arcd39\arcm04\arcs  &   10 &    67 &   139& 49,50\\ 
AS\_034 & 0\uph54\upm42\zdot\ups6 &  -37\arcd43\arcm06\arcs  &   25 &   110 &   230& 53A,53B,53C\\ 
AS\_035 & 0\uph54\upm42\zdot\ups7 &  -37\arcd40\arcm15\arcs  &    6 &    42 &    88& 51\\ 
AS\_036 & 0\uph54\upm42\zdot\ups7 &  -37\arcd41\arcm21\arcs  &   20 &   110 &   229& \\ 
AS\_037 & 0\uph54\upm42\zdot\ups8 &  -37\arcd40\arcm03\arcs  &   12 &    79 &   165& 51,52\\ 
AS\_038 & 0\uph54\upm43\zdot\ups2 &  -37\arcd43\arcm41\arcs  &    6 &    79 &   165& \\ 
AS\_039 & 0\uph54\upm43\zdot\ups3 &  -37\arcd39\arcm12\arcs  &    7 &    53 &   110& \\ 
AS\_040 & 0\uph54\upm43\zdot\ups6 &  -37\arcd45\arcm33\arcs  &   16 &   108 &   225& \\ 
AS\_041 & 0\uph54\upm43\zdot\ups6 &  -37\arcd40\arcm41\arcs  &    8 &    74 &   155& \\ 
AS\_042 & 0\uph54\upm43\zdot\ups6 &  -37\arcd38\arcm37\arcs  &    7 &    73 &   153& \\ 
AS\_043 & 0\uph54\upm44\zdot\ups7 &  -37\arcd42\arcm01\arcs  &    6 &    53 &   110& \\ 
AS\_044 & 0\uph54\upm44\zdot\ups9 &  -37\arcd41\arcm06\arcs  &   14 &    95 &   199& 60\\ 
AS\_045 & 0\uph54\upm45\zdot\ups4 &  -37\arcd38\arcm45\arcs  &    8 &    63 &   132& 61,62\\ 
AS\_046 & 0\uph54\upm45\zdot\ups9 &  -37\arcd40\arcm26\arcs  &   57 &   221 &   462& 56,64\\ 
AS\_047 & 0\uph54\upm46\zdot\ups4 &  -37\arcd42\arcm12\arcs  &   13 &    92 &   193& \\ 
AS\_048 & 0\uph54\upm47\zdot\ups2 &  -37\arcd38\arcm03\arcs  &   16 &    95 &   199& 66,68\\ 
AS\_049 & 0\uph54\upm48\zdot\ups1 &  -37\arcd43\arcm06\arcs  &   11 &    78 &   162& \\ 
AS\_050 & 0\uph54\upm49\zdot\ups1 &  -37\arcd39\arcm48\arcs  &    7 &    62 &   129& 70\\ 
AS\_051 & 0\uph54\upm49\zdot\ups5 &  -37\arcd33\arcm25\arcs  &    6 &    31 &    64& 72\\ 
AS\_052 & 0\uph54\upm49\zdot\ups6 &  -37\arcd38\arcm48\arcs  &  101 &   323 &   675& *, I\\ 
AS\_053 & 0\uph54\upm49\zdot\ups8 &  -37\arcd41\arcm01\arcs  &   25 &   101 &   212& 74,80\\
AS\_054 & 0\uph54\upm50\zdot\ups1 &  -37\arcd35\arcm51\arcs  &    6 &    52 &   109& \\ 
\hline
\multicolumn{7}{l}{ Column 7: Arabic and roman numbers correspond to HII regions (Deharveng}\\
\multicolumn{7}{l}{ et al. 1988), and objects studied by Breysacher et al. (1997), respectively.}\\
\multicolumn{7}{l}{* means that more detailed cross-identification is given in Table 3.}\\
\end{tabular}
\end{table}

\setcounter{table}{0}

\begin{table}

\caption[]{Continued}
\begin{tabular}{ccccccc}
\hline
Name  & $\alpha_{2000}$ & $\delta_{2000}$ & N &size&size & cross- \\
      &                 &                 &   &[$\mu rad$]&[pc]& identification\\
\hline
AS\_055 & 0\uph54\upm50\zdot\ups6 &  -37\arcd36\arcm21\arcs  &    6 &    64 &   134& \\
AS\_056 & 0\uph54\upm50\zdot\ups7 &  -37\arcd40\arcm24\arcs  &   55 &   203 &   424& *\\
AS\_057 & 0\uph54\upm51\zdot\ups4 &  -37\arcd39\arcm44\arcs  &    6 &    58 &   121& 83,84\\
AS\_058 & 0\uph54\upm51\zdot\ups4 &  -37\arcd41\arcm43\arcs  &   34 &   180 &   376& *\\
AS\_059 & 0\uph54\upm52\zdot\ups5 &  -37\arcd38\arcm53\arcs  &   14 &   130 &   271& \\
AS\_060 & 0\uph54\upm53\zdot\ups0 &  -37\arcd43\arcm45\arcs  &   22 &   115 &   241&87,88,90,91,93\\ 
AS\_061 & 0\uph54\upm53\zdot\ups2 &  -37\arcd40\arcm08\arcs  &   11 &    65 &   136& \\ 
AS\_062 & 0\uph54\upm54\zdot\ups0 &  -37\arcd35\arcm48\arcs  &    6 &    63 &   132& \\ 
AS\_063 & 0\uph54\upm54\zdot\ups7 &  -37\arcd36\arcm49\arcs  &   23 &   113 &   236& 92,94\\
AS\_064 & 0\uph54\upm55\zdot\ups5 &  -37\arcd45\arcm44\arcs  &   17 &   137 &   287& 97\\ 
AS\_065 & 0\uph54\upm55\zdot\ups9 &  -37\arcd40\arcm13\arcs  &    9 &    23 &    48& 96\\ 
AS\_066 & 0\uph54\upm56\zdot\ups1 &  -37\arcd44\arcm33\arcs  &    8 &    72 &   150& \\ 
AS\_067 & 0\uph54\upm56\zdot\ups5 &  -37\arcd41\arcm11\arcs  &   13 &    81 &   169& 100\\ 
AS\_068 & 0\uph54\upm57\zdot\ups1 &  -37\arcd40\arcm35\arcs  &    9 &    82 &   172& 98\\ 
AS\_069 & 0\uph54\upm57\zdot\ups2 &  -37\arcd41\arcm34\arcs  &    8 &    68 &   141& \\ 
AS\_070 & 0\uph54\upm57\zdot\ups8 &  -37\arcd45\arcm43\arcs  &    6 &    36 &    75& \\ 
AS\_071 & 0\uph54\upm57\zdot\ups9 &  -37\arcd42\arcm32\arcs  &   10 &   110 &   229& 103,104\\
AS\_072 & 0\uph54\upm58\zdot\ups3 &  -37\arcd41\arcm11\arcs  &    6 &    65 &   136& 105\\ 
AS\_073 & 0\uph54\upm58\zdot\ups8 &  -37\arcd44\arcm19\arcs  &    9 &    85 &   177& 108\\ 
AS\_074 & 0\uph54\upm59\zdot\ups0 &  -37\arcd36\arcm34\arcs  &   12 &    53 &   111& \\ 
AS\_075 & 0\uph54\upm59\zdot\ups1 &  -37\arcd43\arcm20\arcs  &    9 &    84 &   176& \\ 
AS\_076 & 0\uph55\upm00\zdot\ups2 &  -37\arcd40\arcm41\arcs  &   12 &    50 &   104& 109\\ 
AS\_077 & 0\uph55\upm01\zdot\ups2 &  -37\arcd39\arcm29\arcs  &    6 &    62 &   129& \\ 
AS\_078 & 0\uph55\upm01\zdot\ups7 &  -37\arcd42\arcm22\arcs  &    6 &    28 &    58& \\ 
AS\_079 & 0\uph55\upm01\zdot\ups9 &  -37\arcd41\arcm49\arcs  &    6 &    70 &   146& \\ 
AS\_080 & 0\uph55\upm02\zdot\ups1 &  -37\arcd36\arcm53\arcs  &    8 &    48 &   100& \\ 
AS\_081 & 0\uph55\upm02\zdot\ups3 &  -37\arcd44\arcm26\arcs  &   12 &    65 &   136& 113,116\\ 
AS\_082 & 0\uph55\upm02\zdot\ups6 &  -37\arcd38\arcm25\arcs  &   40 &   139 &   291& 115\\ 
AS\_083 & 0\uph55\upm02\zdot\ups9 &  -37\arcd40\arcm04\arcs  &   11 &    53 &   111& \\ 
AS\_084 & 0\uph55\upm03\zdot\ups1 &  -37\arcd42\arcm47\arcs  &   62 &   227 &   475& 118A, III\\ 
AS\_085 & 0\uph55\upm03\zdot\ups3 &  -37\arcd41\arcm21\arcs  &   13 &   121 &   252& 123\\ 
AS\_086 & 0\uph55\upm03\zdot\ups5 &  -37\arcd43\arcm23\arcs  &   30 &   119 &   248& 119A,119B, II\\ 
AS\_087 & 0\uph55\upm03\zdot\ups8 &  -37\arcd40\arcm29\arcs  &    6 &    54 &   112& \\ 
AS\_088 & 0\uph55\upm04\zdot\ups2 &  -37\arcd39\arcm26\arcs  &   16 &   133 &   279& 120\\ 
AS\_089 & 0\uph55\upm04\zdot\ups2 &  -37\arcd41\arcm14\arcs  &    6 &    33 &    70& \\ 
AS\_090 & 0\uph55\upm04\zdot\ups5 &  -37\arcd40\arcm53\arcs  &   16 &    67 &   139& 122\\ 
AS\_091 & 0\uph55\upm05\zdot\ups0 &  -37\arcd41\arcm44\arcs  &    7 &    78 &   163& 123\\ 
AS\_092 & 0\uph55\upm05\zdot\ups3 &  -37\arcd41\arcm26\arcs  &    8 &    49 &   103& 123,124\\ 
AS\_093 & 0\uph55\upm05\zdot\ups9 &  -37\arcd43\arcm14\arcs  &    9 &    99 &   207& 119B\\ 
AS\_094 & 0\uph55\upm06\zdot\ups9 &  -37\arcd41\arcm05\arcs  &   14 &    80 &   168& 126\\ 
AS\_095 & 0\uph55\upm07\zdot\ups6 &  -37\arcd42\arcm58\arcs  &    7 &    55 &   114& \\ 
AS\_096 & 0\uph55\upm08\zdot\ups4 &  -37\arcd40\arcm21\arcs  &   33 &   187 &   392& 130\\ 
AS\_097 & 0\uph55\upm08\zdot\ups7 &  -37\arcd41\arcm58\arcs  &    7 &    80 &   168& 127\\ 
AS\_098 & 0\uph55\upm08\zdot\ups8 &  -37\arcd40\arcm50\arcs  &    6 &    41 &    86& 130\\ 
AS\_099 & 0\uph55\upm08\zdot\ups8 &  -37\arcd39\arcm33\arcs  &   13 &    71 &   148& 129\\ 
AS\_100 & 0\uph55\upm09\zdot\ups3 &  -37\arcd47\arcm56\arcs  &   10 &    45 &    95& 131,133\\ 
AS\_101 & 0\uph55\upm12\zdot\ups6 &  -37\arcd44\arcm12\arcs  &   33 &   177 &   369& *\\ 
AS\_102 & 0\uph55\upm12\zdot\ups7 &  -37\arcd41\arcm31\arcs  &   63 &   174 &   363& *\\ 
AS\_103 & 0\uph55\upm13\zdot\ups4 &  -37\arcd37\arcm47\arcs  &    6 &    36 &    75& \\ 
AS\_104 & 0\uph55\upm14\zdot\ups2 &  -37\arcd40\arcm16\arcs  &    6 &    49 &   103& \\ 
AS\_105 & 0\uph55\upm14\zdot\ups2 &  -37\arcd41\arcm16\arcs  &    6 &    49 &   102& \\ 
AS\_106 & 0\uph55\upm15\zdot\ups1 &  -37\arcd42\arcm05\arcs  &    6 &    37 &    77& \\ 
\hline
\multicolumn{7}{l}{ Column 7: Arabic and roman numbers correspond to HII regions (Deharveng}\\
\multicolumn{7}{l}{ et al. 1988), and objects studied by Breysacher et al. (1997), respectively.}\\
\multicolumn{7}{l}{* means that more detailed cross-identification is given in Table 3.}
\end{tabular}
\end{table}

\setcounter{table}{0}

\begin{table}

\caption[]{Continued}
\begin{tabular}{ccccccc}
\hline
Name  & $\alpha_{2000}$ & $\delta_{2000}$ & N &size&size & cross- \\
      &                 &                 &   &[$\mu rad$]&[pc]& identification\\
\hline
AS\_107 & 0\uph55\upm15\zdot\ups6 &  -37\arcd44\arcm30\arcs  &   15 &   101 &   210& 140,141,142\\ 
AS\_108 & 0\uph55\upm19\zdot\ups2 &  -37\arcd44\arcm09\arcs  &   10 &    83 &   174& \\ 
AS\_109 & 0\uph55\upm20\zdot\ups7 &  -37\arcd43\arcm49\arcs  &   17 &   141 &   294& 145,146\\ 
AS\_110 & 0\uph55\upm24\zdot\ups5 &  -37\arcd39\arcm32\arcs  &   11 &    66 &   139& 147\\ 
AS\_111 & 0\uph55\upm30\zdot\ups2 &  -37\arcd39\arcm11\arcs  &   12 &    93 &   194& 153,154\\ 
AS\_112 & 0\uph55\upm30\zdot\ups9 &  -37\arcd36\arcm05\arcs  &    6 &    11 &    24& \\ 
AS\_113 & 0\uph55\upm33\zdot\ups4 &  -37\arcd43\arcm16\arcs  &   11 &    66 &   137& 159\\ 
AS\_114 & 0\uph55\upm33\zdot\ups8 &  -37\arcd41\arcm15\arcs  &   11 &    63 &   131& \\ 
AS\_115 & 0\uph55\upm36\zdot\ups0 &  -37\arcd41\arcm50\arcs  &    9 &    62 &   130& 161\\ 
AS\_116 & 0\uph55\upm36\zdot\ups2 &  -37\arcd41\arcm29\arcs  &   10 &    90 &   189& \\ 
AS\_117 & 0\uph55\upm38\zdot\ups2 &  -37\arcd44\arcm03\arcs  &    8 &    66 &   138& \\ 
\hline
\multicolumn{7}{l}{ Column 7: Arabic and roman numbers correspond to HII regions (Deharveng}\\
\multicolumn{7}{l}{ et al. 1988), and objects studied by Breysacher et al. (1997), respectively.}\\
\end{tabular}
\end{table}

\setcounter{table}{2}

\begin{table}

\caption[]{OB association candidates in the NGC 300 found
in the regions of potential stellar-complexes}

\begin{tabular}{ccccccc}
\hline
Name  & $\alpha_{2000}$ & $\delta_{2000}$ & N & size & size & cross- \\
      &                 &                 &   &[$\mu rad$]      &[pc]&identification\\
\hline
AS\_002a & 0\uph54\upm16\zdot\ups3& -37\arcd34\arcm57\arcs &  19&  55 & 114&6\\
AS\_002b & 0\uph54\upm17\zdot\ups8& -37\arcd35\arcm07\arcs &  6&  26 & 53&9\\
AS\_002c & 0\uph54\upm16\zdot\ups9 & -37\arcd35\arcm18\arcs& 24& 131 & 273&7,8\\
AS\_002d & 0\uph54\upm17\zdot\ups2 & -37\arcd34\arcm46\arcs & 6 & 26 & 53&\\ \hline
AS\_009a & 0\uph54\upm25\zdot\ups3 &-37\arcd39\arcm46\arcs &20 &24 & 130& 15,14\\
AS\_009b & 0\uph54\upm25\zdot\ups6 &-37\arcd39\arcm15\arcs &14 & 15 & 32& 17 \\ \hline
AS\_014a & 0\uph54\upm28\zdot\ups5 & -37\arcd41\arcm36\arcs & 17 & 58 & 121&24 \\
AS\_014b & 0\uph54\upm29\zdot\ups0 &-37\arcd41\arcm17\arcs & 7 &  45 & 93& \\ \hline
AS\_016a & 0\uph54\upm30\zdot\ups0 &-37\arcd37\arcm52\arcs & 9 & 62 & 129& \\
AS\_016b & 0\uph54\upm31\zdot\ups3 &-37\arcd37\arcm54\arcs & 25 & 108 & 225& 30,31\\ \hline
AS\_018a & 0\uph54\upm30\zdot\ups6 &-37\arcd38\arcm39\arcs & 8 & 49 & 101&\\
AS\_018b & 0\uph54\upm32\zdot\ups1 &-37\arcd38\arcm43\arcs & 10 & 54 & 112&32,34\\ \hline
AS\_026a & 0\uph54\upm38\zdot\ups4 &-37\arcd42\arcm41\arcs & 8 & 26 & 53& 40 \\
AS\_026b & 0\uph54\upm40\zdot\ups2 &-37\arcd42\arcm50\arcs & 15 & 58 & 121& \\ \hline
AS\_027a & 0\uph54\upm39\zdot\ups0 &-37\arcd40\arcm21\arcs & 20 & 74 & 153&\\
AS\_027b &  0\uph54\upm40\zdot\ups6& -37\arcd40\arcm20\arcs & 7 & 32 & 66 &\\ \hline
AS\_029a & 0\uph54\upm41\zdot\ups8& -37\arcd40\arcm59\arcs & 15 & 70 & 145&\\
AS\_029b & 0\uph54\upm40\zdot\ups3& -37\arcd40\arcm54\arcs &  8 & 32 & 66&45\\
AS\_029c & 0\uph54\upm41\zdot\ups1& -37\arcd40\arcm47\arcs &  6 & 45 & 93&45\\ \hline
AS\_046a & 0\uph54\upm46\zdot\ups2& -37\arcd40\arcm14\arcs & 8 & 56 & 115& 64\\
AS\_046b & 0\uph54\upm46\zdot\ups1& -37\arcd40\arcm31\arcs & 19 & 114 & 237&64\\ \hline
AS\_052a & 0\uph54\upm50\zdot\ups2& -37\arcd38\arcm24\arcs & 9 & 54 & 112& 77, I\\
AS\_052b & 0\uph54\upm49\zdot\ups7& -37\arcd38\arcm38\arcs & 17 & 64 & 133& I\\
AS\_052c & 0\uph54\upm49\zdot\ups5& -37\arcd38\arcm51\arcs & 6 & 35 & 73&I\\
AS\_052d & 0\uph54\upm51\zdot\ups0& -37\arcd38\arcm26\arcs & 8 & 51 & 106&79, I\\
AS\_052e & 0\uph54\upm49\zdot\ups8& -37\arcd39\arcm20\arcs & 18 & 70 & 145& 75,71\\ \hline
AS\_056a  & 0\uph54\upm49\zdot\ups6& -37\arcd40\arcm03\arcs & 7 & 43 & 90&76B\\ 
AS\_056b  & 0\uph54\upm49\zdot\ups9& -37\arcd40\arcm19\arcs & 39 & 116 & 241&76B \\ 
AS\_056c  & 0\uph54\upm50\zdot\ups6& -37\arcd40\arcm27\arcs & 9  & 43 & 89& 76A,76C\\ \hline
AS\_058a  & 0\uph54\upm50\zdot\ups7& -37\arcd41\arcm46\arcs & 10 & 74 & 154& 81\\ 
AS\_058b  & 0\uph54\upm51\zdot\ups8& -37\arcd41\arcm49\arcs & 13 & 67 & 139& 81,85\\ \hline
AS\_101a  & 0\uph55\upm11\zdot\ups5& -37\arcd44\arcm06\arcs & 7 & 38 & 80&135\\
AS\_101b  & 0\uph55\upm12\zdot\ups7& -37\arcd44\arcm15\arcs & 13 & 68 & 140& 135,139\\ \hline
AS\_102a  & 0\uph55\upm12\zdot\ups0& -37\arcd41\arcm28\arcs & 34 &116  &  241& 137B,137D\\
AS\_102b  & 0\uph55\upm12\zdot\ups8& -37\arcd41\arcm41\arcs & 20  &67 & 138& 137A,137B \\
AS\_102c  & 0\uph55\upm13\zdot\ups7& -37\arcd41\arcm40\arcs & 13 &43 & 89& 137C \\ \hline 
\multicolumn{7}{l}{ Column 7: Arabic and roman numbers correspond to HII regions (Deharveng}\\
\multicolumn{7}{l}{ et al. 1988), and objects studied by Breysacher et al. (1997), respectively.}\\
\end{tabular}
\end{table}

\vskip0.7cm

\centerline{\Large \bf Figure Captions}

\vskip0.5cm

\noindent
Fig.~1. CMD diagram for about 14000 stars observed in the field of NGC 300. The
rectangular region contains
blue stars used for searching for OB associations.

\vskip0.3cm

\noindent
Fig.~2. {Behaviour of the function $f_{\rm p} (d)$ for a set of p 
(minimum number of stars in an association)   values.

\vskip0.3cm

\noindent
Fig.~3. The percantage of expected spurious detections of associations,
 versus  p, the minimum number of stars in an association.

\vskip0.3cm

\noindent
Fig.~4. Spatial distribution of detected OB associations (filled circles)
over the disc of NGC 300. Dots are individual blue stars.
The center of the figure corresponds to  $\alpha_{2000}$
= 0\uph54\upm53\zdot\ups4,  $\delta_{2000}$ = -37\arcd41\arcm03\arcs.

\vskip0.3cm

\noindent
Fig.~5. Size distribution of associations in NGC 300.


\begin{thebibliography}{}
\bibitem[1991]{bati} Battinelli, P. 1991, A\&A 244, 69
\bibitem[1996]{bati1} Battinelli, P., Efremov, Y., \& Magnier , A.E 
 1996, A\&A 314, 51
\bibitem[1997]{brey} Breysacher, J., Azzopardi, M., Testor, G., 
\&  Muratorio, G. 1997, A\&A 326, 976
\bibitem[1998]{dehar} Deharveng, L., Caplan, J., Legueux, J., et al. 1988 , A\&AS  73, 407
\bibitem[1987]{efre} Efremov, Y.N., Ivanov, G.R., Nokolov, N.S. 1987, Ap\&SS, 135, 119
\bibitem[1992]{free} Freedman, W.L., Madore, B.F., Hawley, S.L., et al.  1992, ApJ, 396, 80
\bibitem[1986]{hodge} Hodge, P.W. 1986, in: Luminous Stars and Associations in Galaxies, 
IAU Symp. 116, eds. C.W.H. De Loore, A.J. Willis, P.G. Laskarides, Reidel; 
Dordrecht, 369
\bibitem[1992]{landolt} Landolt, A.U. 1992, AJ, 104, 372
\bibitem[1993]{mag} Magnier, E.A., Battinelli, P., Lewin, W.H.G., et al.
1993 A\&A, 278, 36
\bibitem[1997]{udalski} Udalski, A., Kubiak, M., \& Szyma\'nski, M. 1997,  Acta Astron., 
47, 319
\bibitem[1992]{van}  van den Bergh, S. 1992, P.A.S.P. 104, 861 
\bibitem[1991]{wil1} Wilson, C.D. 1991, AJ, 101, 1663
\bibitem[1992]{wil2} Wilson, C.D. 1992, ApJ, 386, L29
\end{thebibliography}
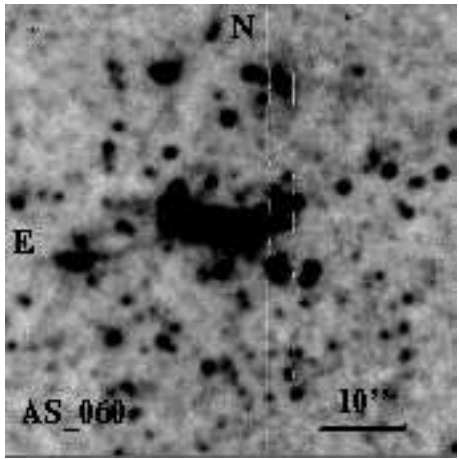
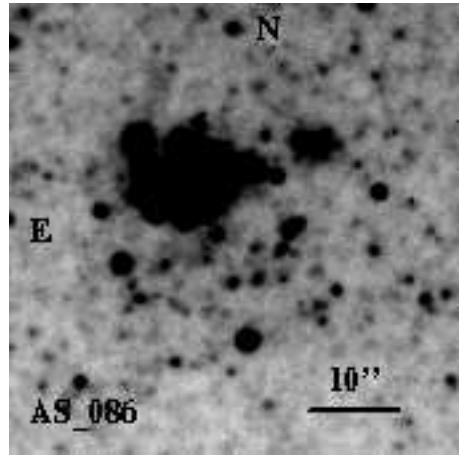
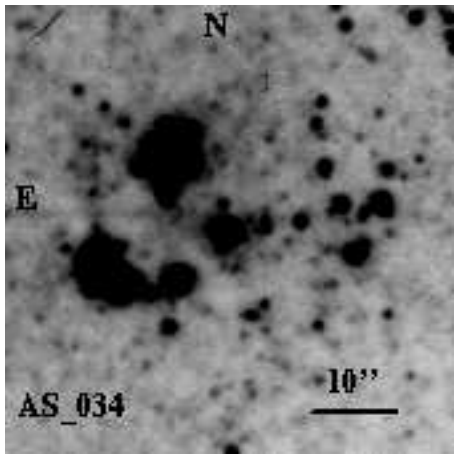
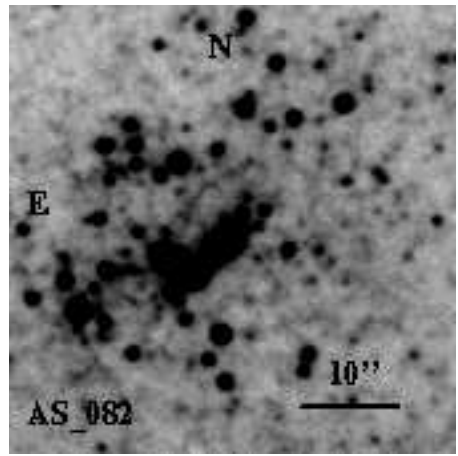
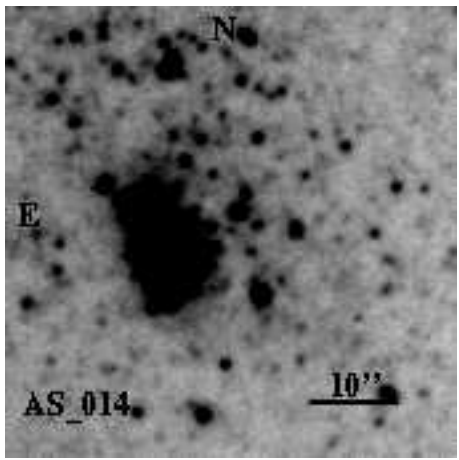
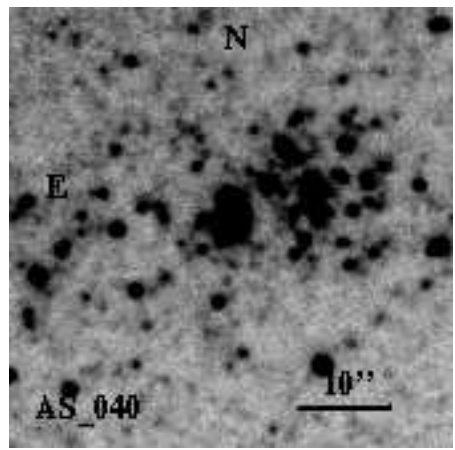
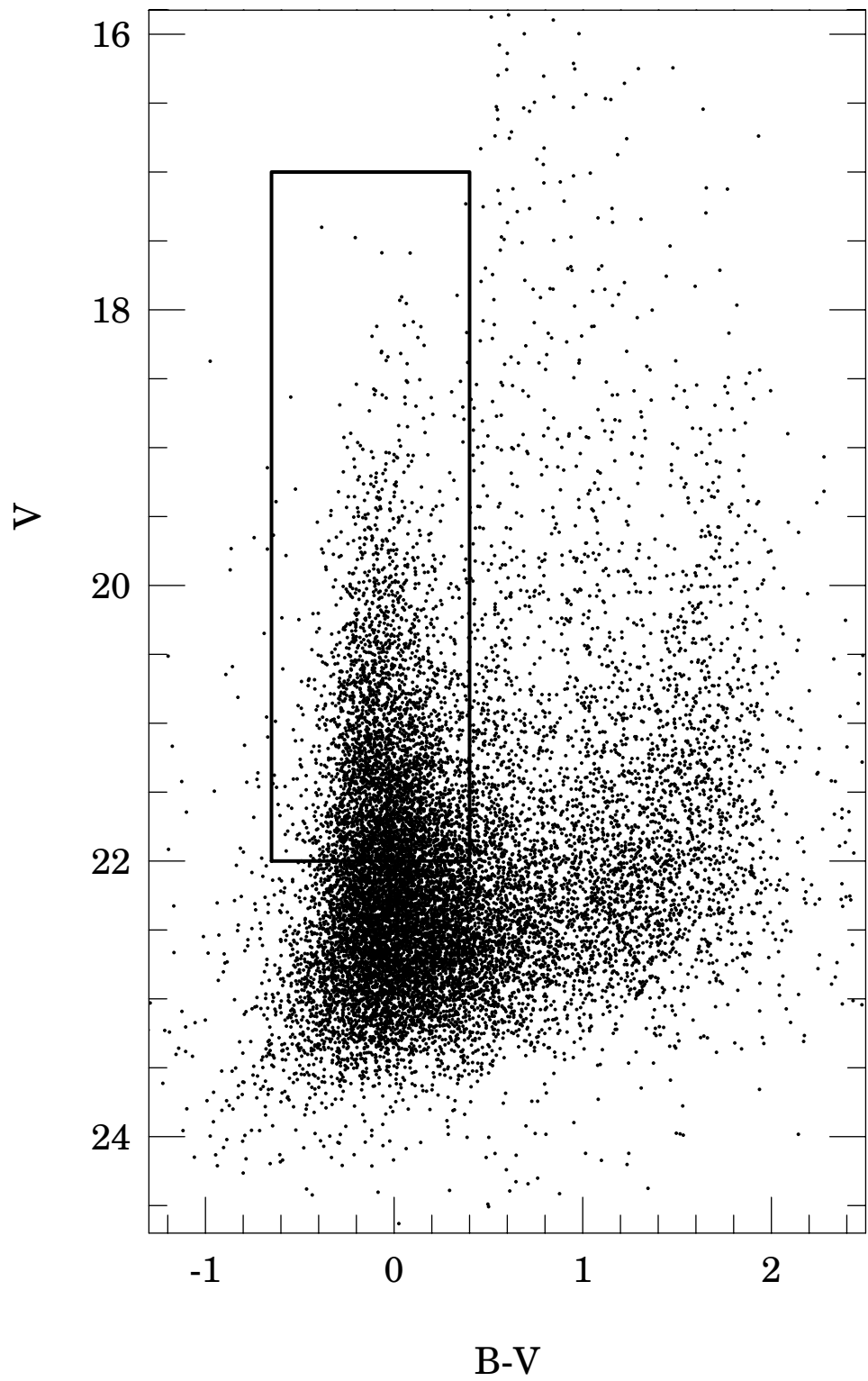
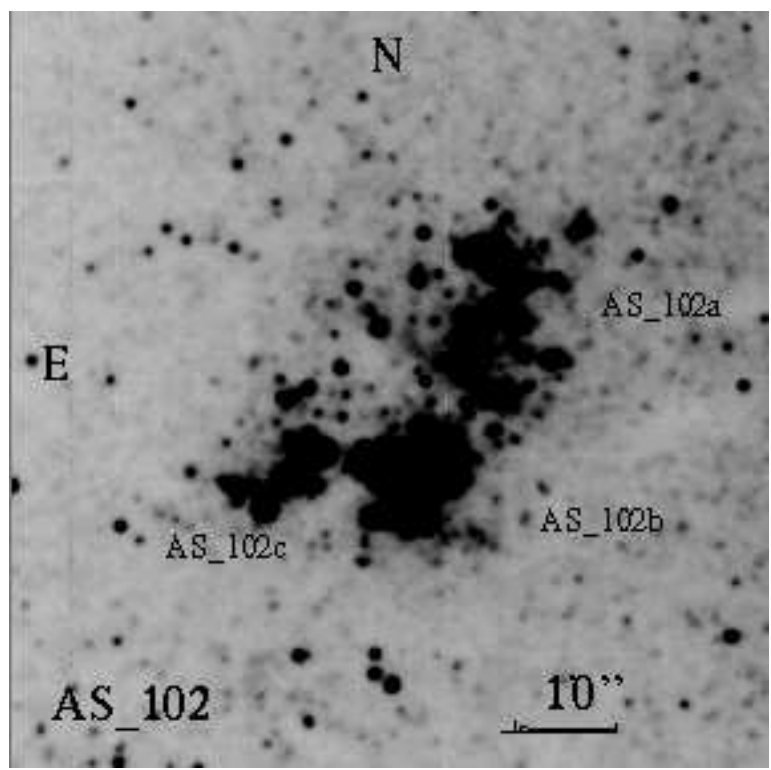
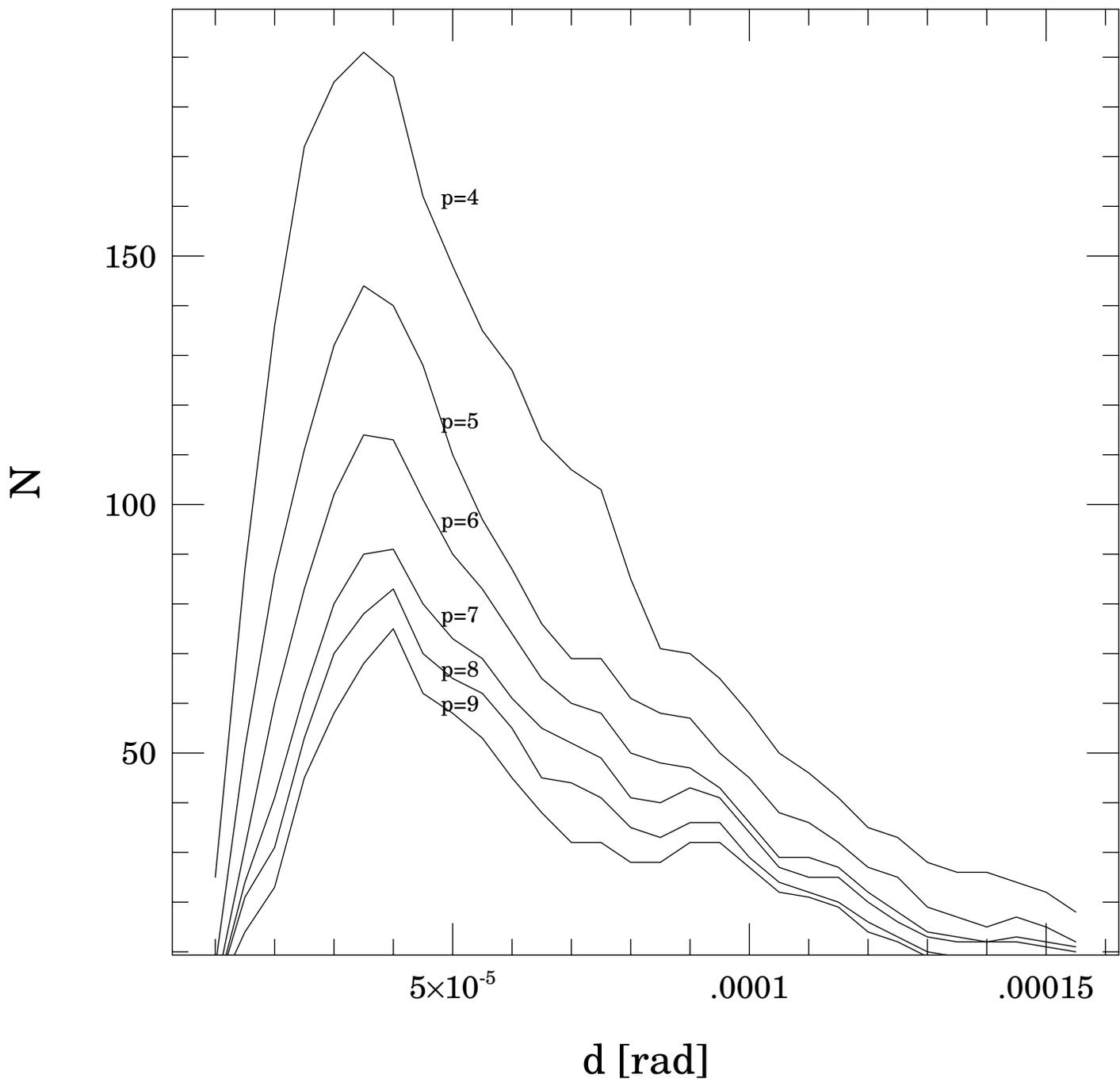
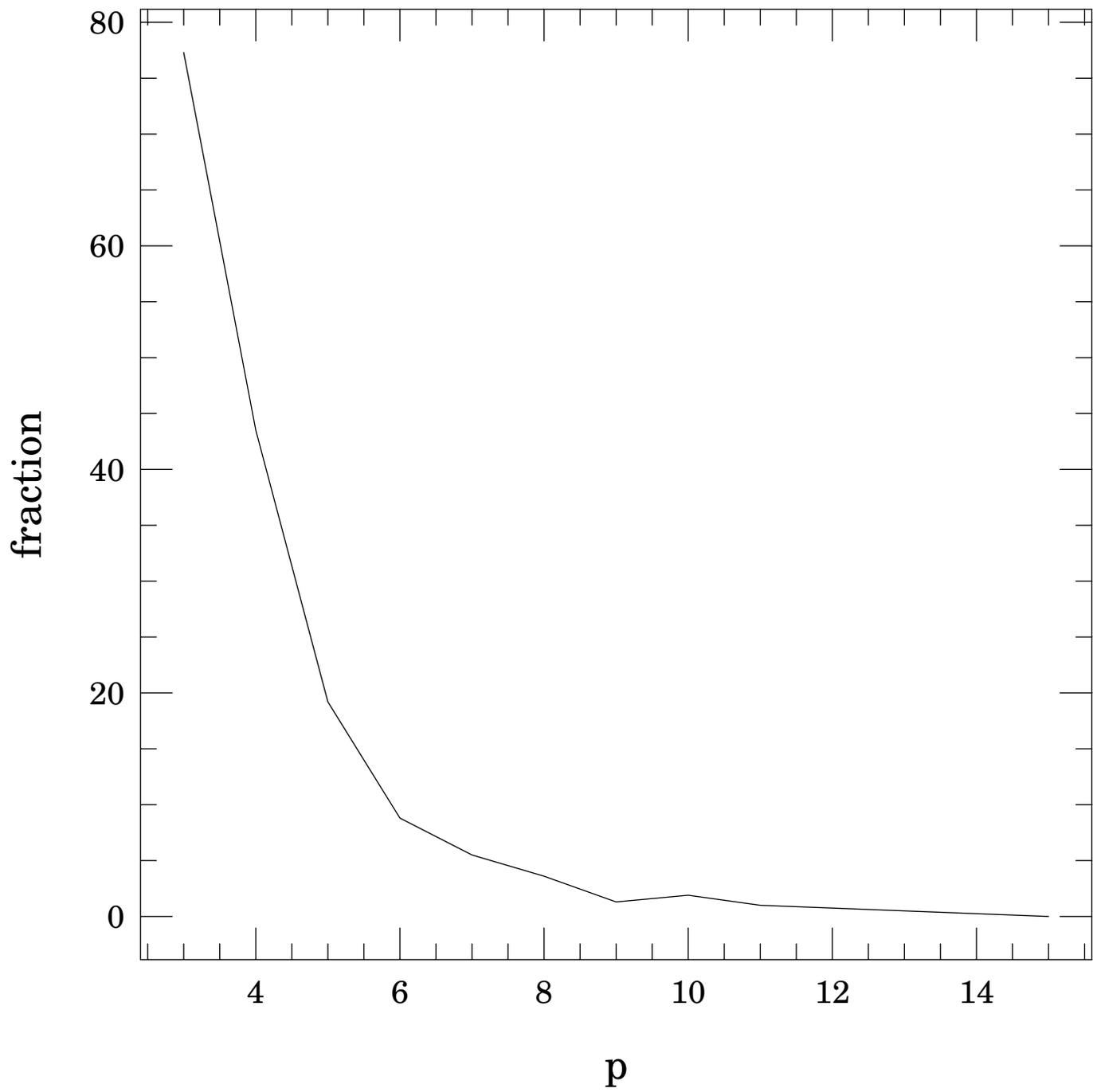
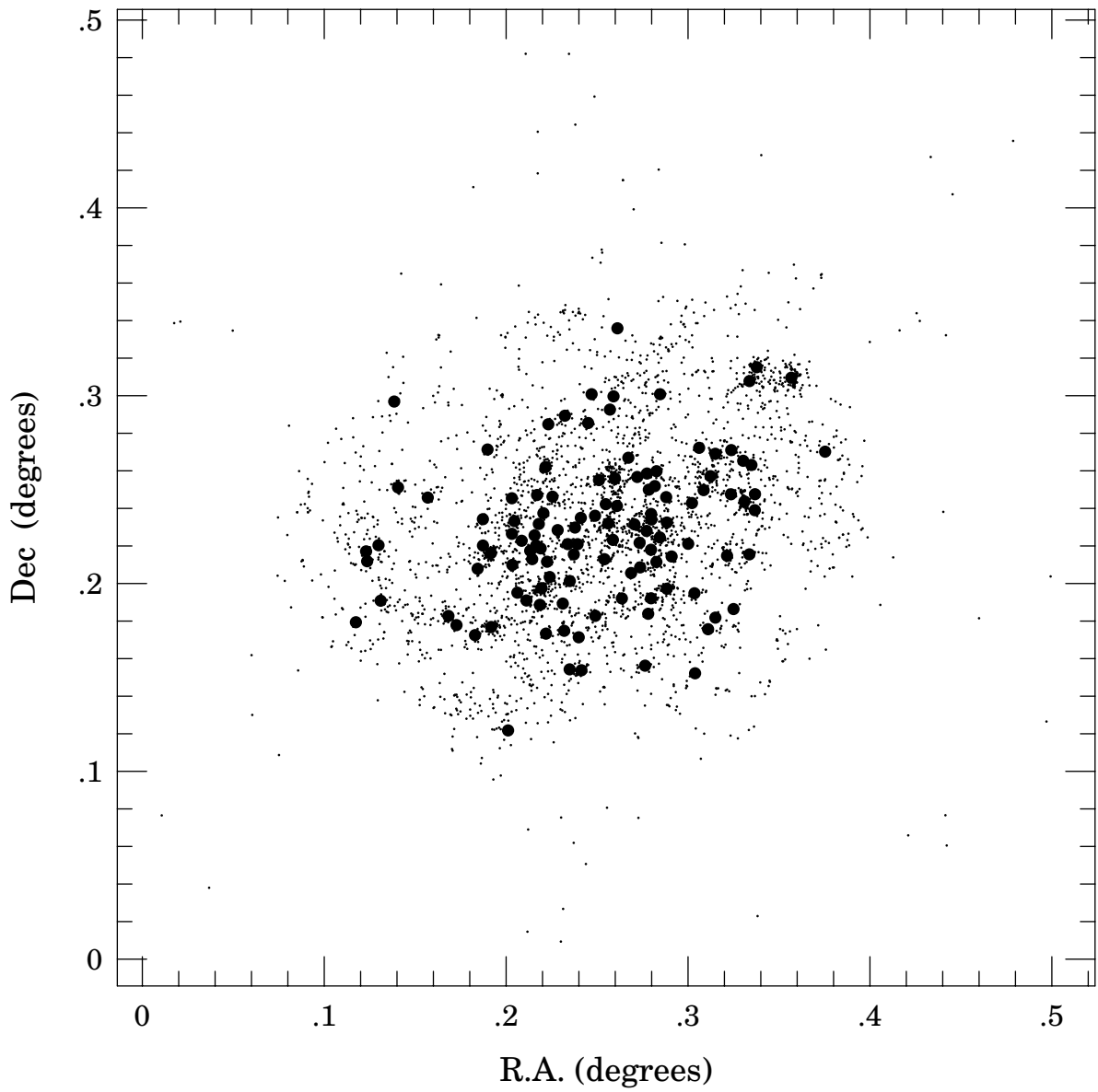
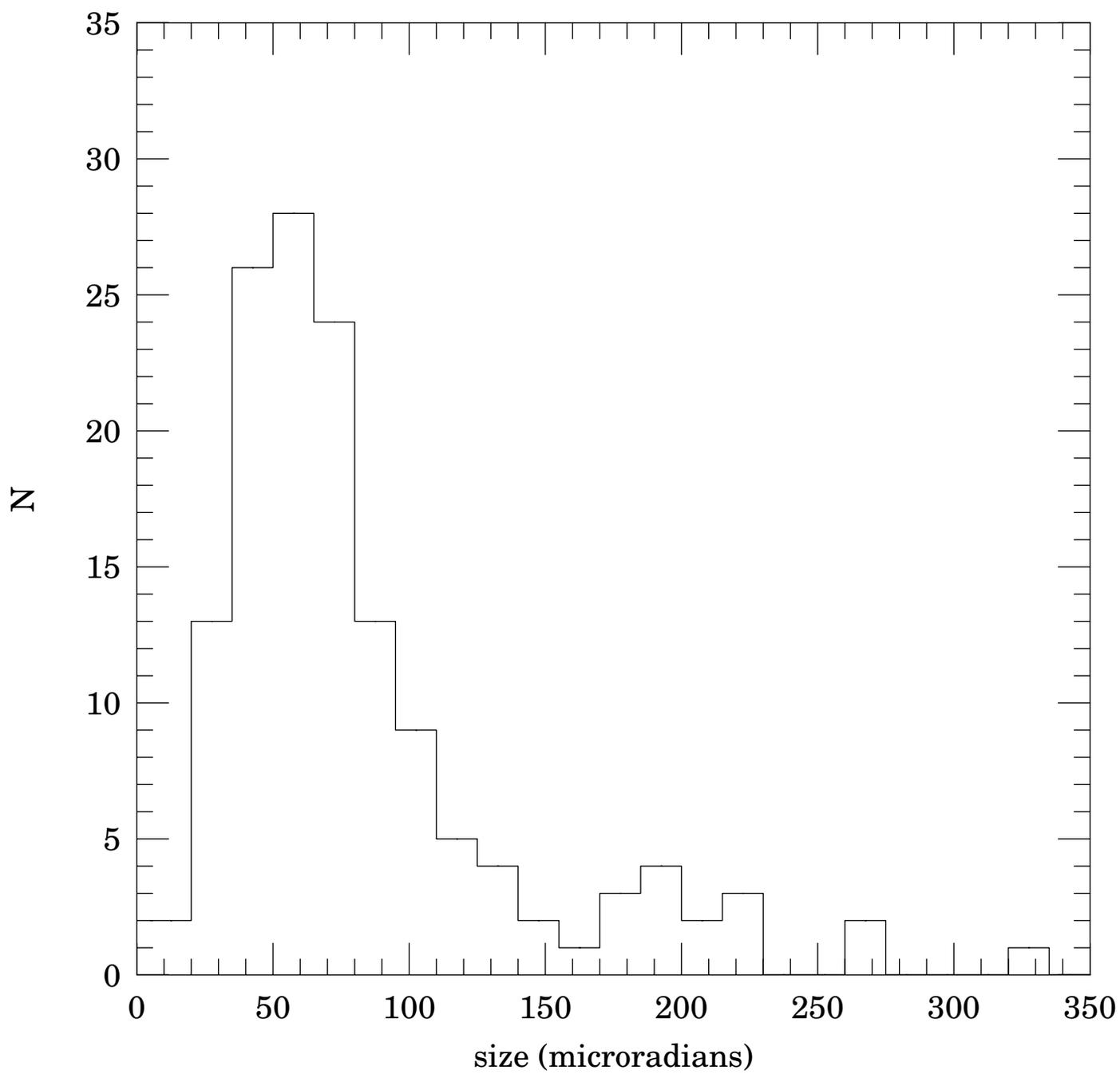
\end{document}